# DOES RANDOMNESS IN MULTINOMIAL MEASURES IMPLY NEGATIVE DIMENSIONS?

WEI-XING ZHOU

*Institute of Geophysics and Planetary Physics, University of California, Los Angeles, CA 90095, USA*
*Institute of Clean Coal Technology, East China University of Science and Technology, Box 272, Shanghai, 200237, PR China*
*E-mail: wxzhou@moho.ess.ucla.edu*

ZUN-HONG YU

*Institute of Clean Coal Technology, East China University of Science and Technology, Box 272, Shanghai, 200237, PR China*
*E-mail: zhyu@ecust.edu.cn*

Negative, or latent, dimensions have always attracted a strong interest since their discovery. When randomness is introduced in multifractals, the sample-to-sample fluctuations of multifractal spectra emerge inevitably, which has motivated various studies in this field. In this work, we study a class of multinomial measures and argue the asymptotic behaviors of the multifractal function $f(\alpha)$ as $q \to \pm\infty$. The so-called latent dimensions condition (LDC) is presented which states that latent dimensions may be absent in discrete random multinomial measures. In order to clarify the discovery, several examples are illustrated.

## 1 Introduction

Multifractal analysis [24,29,25,23,28] has been used as a very useful technique in the analysis of singular measures, which gives rise to many scientific fields, both in natural aspects and social facets, say, the fully developed turbulence [35,3,49,44,22,5,7,34], diffusion-limited aggregations [41,33,6,52,4] and chaotic dynamical systems [27,26]. Generally, there are two classes of multifractal measures: one is deterministic self-similar, and the other is statistically self-similar [19,1] upon which most attentions have focused in the latest decade. It is clear that, most of the physical processes are random, which has led to the sample-to-sample fluctuations of the multifractal function by an amount greater than the error bars on any one sample would indicate [16] and, in general, the negative dimensions arise in random multifractals.

The randomness of multifractals at least arises in two situations where negative dimensions may emerge. First, one may obtain a multifractal constructed by a multiplicative cascade that is inherently probabilistic and the negative dimensions, if exist, describe the rarely occurring events. Second, one may have to investigate the experiment in a probabilistic view. For instance, one-dimensional cuts of a deterministic measure carried by a deterministic Sierpinski sponge will inevitably introduce randomness and may be regarded as random samples of a population. Also, when one performs measurements upon the turbulence with dual PDA [55] or hot-wire anemometry [44], one-dimensional cuts are obtained, while two-dimensional cuts are obtained to describe the scalar fields when CCD and LIF technique are utilized [53,49,51]. Such measurements unavoidably introduce randomness. In such systems, negative dimensions often appear and are used to characterize the sample-to-sample fluctuation of multifractal spectrum.

The physical meaning of negative dimensions was discussed foremostly by Cates and Witten [10]. However, it was Mandelbrot who recognized the essential importance of negative dimensions for both physical phenomena and mathematical realizations and then studied them systematically based on turbulence and multinomial measures [36,37,38,39,40]. Then, Evertsz and Mandelbrot applied Chernoff's theorem [17] on large deviations to computer the tail distribution of the coarse Hölder exponent of a randomly picked interval of the multinomial measure of finite discrete multipliers [18] and pointed out that more general Cramèr type large deviation theorems provide a full justification of the so-called thermodynamic formalism of multifractals based on the Legendre transforms [18,22]. Also, with a simple but cogitative analytical example, Chhabra and Sreenivasan [15] tried to propose a theory of negative dimensions. It is natural that different methods have been developed to extract the $f(\alpha)$ function, such as the method of moments [28,26], histograms method [43], method of supersampling [38], canonical method [13,14], multiplier method [15,16,29,30], and gliding box method [12].

However, these points are intuitive heuristic interpretations rather than mathematical descriptions. It is well known that multifractal formalism involves decomposition fractal measure into interwoven fractal sets each of which is characterized by its singularity strength [11,8,45]. When concerning with statistically self-similar measures, Falconer [19] proved that there is an open interval $A = (\alpha_{\min}, \alpha_{\max})$ such that, for all $\alpha \in A$, $\dim F_\alpha = f(\alpha)$ with probability one if $f(\alpha) \geq 0$. We postpone the introduction of the rigorous definitions of notions. Although he said nothing in his theorem about the situation when $f(\alpha) < 0$, he gave an example where negative dimensions do appear. In a more recent work, Barral [2] improved the previous result and claimed that, with probability one, for all $\alpha \in A$, $\dim F_\alpha = f(\alpha)$. Moreover, Olsen [48] has made rigorous some aspects of Mandelbrot's intuition that negative dimensions may be explained geometrically by considering cuts of higher dimensional multifractals. The multifractal slice theorem [46,47,48,49] presents a mathematical interpretation of negative dimensions and is the basis of experimental measurement of lower dimensional cuts when the ensemble measurement is difficult to be carried out [49,44,55,53]. These works are the basis of discussing negative dimensions in this paper.

## 2 Random multinomial measure

We define a construction rule matrix **CRM** by

$$\mathbf{CMR} = \left\{ \begin{array}{ccc|ccc|c} r(1,1) & \cdots & r(1,b) & m(1,1) & \cdots & m(1,b) & p(1) \\ \vdots & & \vdots & \vdots & & \vdots & \vdots \\ r(n,1) & \cdots & r(n,b) & m(n,1) & \cdots & m(n,b) & p(n) \end{array} \right\} \quad (1)$$

Then each division at certain generation for each sample of the assemble measures corresponds to a row of **CRM**. Based on Falconer's result [19], we find that, an alternative matrix of the construction rules validates:

$$\mathbf{CRM} = \left\{ \begin{array}{ccccc} m_1 & \cdots & m_{(i-1)b+j} & \cdots & m_{nb} \\ r_1 & \cdots & r_{(i-1)b+j} & \cdots & r_{nb} \\ p_1 & \cdots & p_{(i-1)b+j} & \cdots & p_{nb} \end{array} \right\} \quad (2)$$

Here $m_{(i-1)b+j} = m(i,j)$, $r_{(i-1)b+j} = r(i,j)$ and $p_{(i-1)b+j} = p(i)$. Note that $\sum_{i=1}^{nb} p_i = b$. Moreover, if there are $i$ and $j$ satisfying $m_i = m_j$ and $r_i = r_j$ then we can withdraw the $j$ th column of **CRM** and replace $p_i$ with $p_i + p_j$. Thus performing this procedure simplifies the construction rule matrix to a $3 \times N$ matrix with $N \leq nb$.

Let $\tau(q)$ be the root to the equation $\Gamma(q,\tau) = \sum_1^N p_i m_i^q / r_i^\tau = 1$. Since $\tau'(q) > 0$ and $q'(\tau) > 0$, for fixed $q$, there is a unique root $\tau(q)$. We denote, for future use, that

$$w_i(q) = m_i^q / r_i^\tau. \quad (3)$$

Hence,

$$\sum_{i=1}^N p_i w_i(q) = 1. \quad (4)$$

Define a function $f(\alpha)$, which is related to the moment exponent $\tau(q)$ by Legendre transform and inverse Legendre transform

$$\alpha = \tau'(q) \text{ and } f(\alpha) = q\alpha - \tau(q). \quad (5)$$

It is now known [2] that the Hausdorff dimension $\dim F_\alpha$ of $F_\alpha$ is identical to $f(\alpha)$ with probability one for all $\alpha \in \left( \min_i \{\ln m_i / \ln r_i\}, \max_i \{\ln m_i / \ln r_i\} \right)$.

A latent dimension refers to $f(\alpha) < 0$ but $\alpha > 0$ and characterizes the emptiness of the corresponding set $F_\alpha = \Phi$. A more desirable definition of negative dimensions might be to say that a random self-similar measure has negative dimensions if the Legendre transform of $\tau(q)$ has negative values, and that this definition of negative dimensions is only possible for self-similar measures, and there is no obvious way to define the notion of negative dimensions for an arbitrary measure.

## 3 Asymptotic behaviors

Let $I = \{1, 2, \cdots, N\}$ be the set of column indexes of **CRM** and $M$ be the union of $\underline{M}$ and $\overline{M}$, where

$$\overline{M} = \left\{ i \in I : \ln m_i / \ln r_i = \max_{j \in I} \{\ln m_i / \ln r_i\} \right\} \tag{6}$$

and

$$\underline{M} = \left\{ i \in I : \ln m_i / \ln r_i = \min_{j \in I} \{\ln m_i / \ln r_i\} \right\}. \tag{7}$$

From the previous subsection, we have

$$\alpha(q) = \sum_{i=1}^{N} p_i w_i(q) \ln m_i \bigg/ \sum_{i=1}^{N} p_i w_i(q) \ln r_i. \tag{8}$$

Hence,

$$w_i'(q) = [\ln m_i / \ln r_i - \alpha(q)] w_i(q) \ln r_i. \tag{9}$$

It is well known, in non-trivial case of random multifractal formalism, that

$$\alpha'(q) < 0, \tag{10}$$

$$\lim_{q \to \infty} \alpha(q) = \ln m_i / \ln r_i \tag{11}$$

for $i \in \underline{M}$, and

$$\lim_{q \to -\infty} \alpha(q) = \ln m_i / \ln r_i \tag{12}$$

for $i \in \overline{M}$. Therefore, according to Eq.(9), we find for $i \in \overline{M}$

$$w_i'(q) < 0 \tag{13}$$

and for $i \in \underline{M}$

$$w_i'(q) > 0. \tag{14}$$

In addition, we find for all $i \in I - M$ that there is a unique stagnation point $q_{\text{stag}}$ of $w_i(q)$ and that

$$w_i'(q) > 0 \text{ if } q < q_{\text{stag}} \tag{15}$$

and

$$w_i'(q) < 0 \text{ if } q > q_{\text{stag}}. \tag{16}$$

Since $w_i(q)$ is bounded for all $i \in I$, the limits $\lim_{q \to \pm\infty} w_i(q)$ exist and are finite, which we denote as $w_i(\pm\infty)$ hereafter. Moreover, $w_i(-\infty)$ for $i \in \overline{M}$ and $w_i(\infty)$ for $i \in \underline{M}$ are positive. By taking limit $q \to -\infty$ of identity

$$\ln r_i \ln w_j(q) = q(\ln m_j \ln r_i - \ln m_i \ln r_j) + \ln r_j \ln w_i(q) \tag{17}$$

and using

$$\ln m_j \ln r_i - \ln m_i \ln r_j \neq 0 \tag{18}$$

for $i \in \overline{M}$ and $j \in I - \overline{M}$, it is easy to see that

$$w_{j \notin \overline{M}}(-\infty) = 0 \tag{19}$$

Similarly,
$$w_{j \notin \underline{M}}(\infty) = 0. \tag{20}$$

If we write
$$\phi_j^i = \ln m_i / \ln m_j, \tag{21}$$

then for $i, j \in \overline{M}$ or $i, j \in \underline{M}$
$$w_i(q) = \left[w_j(q)\right]^{\phi_j^i}. \tag{22}$$

One finds immediately that for arbitrary $j \in \overline{M}$ there is a unique solution $w_j(-\infty)$ to equation
$$\sum_{i \in \overline{M}} p_i \left[w_j(-\infty)\right]^{\phi_j^i} = 1 \tag{23}$$

and that for arbitrary $j \in \underline{M}$ there exists a unique solution $w_j(\infty)$ to equation
$$\sum_{i \in \underline{M}} p_i \left[w_j(-\infty)\right]^{\phi_j^i} = 1. \tag{24}$$

Therefore, it follows that for $i \in \overline{M}$
$$\lim_{q \to -\infty} f[\alpha(q)] = \ln w_i(-\infty) / \ln r_i \tag{25}$$

and that for $i \in \underline{M}$
$$\lim_{q \to \infty} f[\alpha(q)] = \ln w_i(\infty) / \ln r_i. \tag{26}$$

## 4  Discussion and illustrations

### 4.1  Latent Dimensions Condition

Consider function $h(w) = \sum_{k=1}^{K} p_k w^{z_k} - 1$ where $0 < p_k < 1$ and $z_k > 0$. Let us now investigate the solution nature of $h(w) = 0$. Since $h'(w) > 0$, $h(0) < 0$ and $h(w) > 0$ for certain $w > 0$ satisfying $p_k w^{z_k} > 1$ for any $k$, it follows from the intermediate value theorem that the solution $w$ to $h(w) = 0$ exists uniquely. If $\sum_{k=1}^{K} p_k > 1$, then $h(1) > 0$ and hence $0 < w < 1$. If $\sum_{k=1}^{K} p_k = 1$, then $w = 1$. If $\sum_{k=1}^{K} p_k < 1$, then $h(1) < 0$ and hence $w > 1$. These three cases lead to three types of asymptotic behaviors at the either endpoints of the multifractal spectrum.

At the right endpoint of the $f(\alpha)$ curve, there exist three situations according to Eq.(25). If $\sum_{i \in \overline{M}} p_i > 1$, then $0 < w_i(-\infty) < 1$ for all $i \in \overline{M}$ and hence $\lim_{q \to -\infty} f[\alpha(q)] > 0$. If $\sum_{i \in \overline{M}} p_{i_k} = 1$, then $w_i(-\infty) = 1$ for all $i \in \overline{M}$ and hence $\lim_{q \to -\infty} f[\alpha(q)] = 0$. If $\sum_{i \in \overline{M}} p_i < 1$, then $w_i(-\infty) > 1$ for all $i \in \overline{M}$ and hence $\lim_{q \to -\infty} f[\alpha(q)] < 0$. Similarly, at the left endpoint of the $f(\alpha)$ curve, there are three situations according to Eq.(26). If $\sum_{i \in \underline{M}} p_i > 1$, then $0 < w_i(\infty) < 1$ for all $i \in \underline{M}$ and hence $\lim_{q \to \infty} f[\alpha(q)] > 0$. If $\sum_{i \in \underline{M}} p_i = 1$, then $w_i(\infty) = 1$ for all $i \in \underline{M}$ and hence $\lim_{q \to \infty} f[\alpha(q)] = 0$. If $\sum_{i \in \underline{M}} p_i < 1$, then $w_i(\infty) > 1$ for all $i \in \underline{M}$ and hence $\lim_{q \to \infty} f[\alpha(q)] < 0$.

Since $f''(\alpha) < 0$ for non-trivial random multifractals, that is, the $f(\alpha)$ curve is somewhat $\cap$-shaped, latent dimensions exist if and only if $\sum_{i \in \overline{M}} p_i < 1$ or $\sum_{i \in \underline{M}} p_i < 1$, which may be called the Latent Dimensions Condition (LDC).

### 4.2 Crossover from random to determinstic Model Behavior

Consider now a class of random multinomial measures [15] whose **CRM** are represented in the form of Eq.(1) satisfying

$$\{(r(i,k), m(i,k)): k = 1, 2, \cdots, b\} = \{(r(j,k), m(j,k)): k = 1, 2, \cdots, b\} \quad (27)$$

for all $1 \leq i, j \leq n$. Hence $p_k = 1$ for $1 \leq k \leq b$ in the transformed **CRM**. This leads to an identical multifractal function just as in the case of deterministic multifractals. To clarify this point, let us investigate a very simple example. Consider $m \in (0, 0.5)$. Construct the multinomial measure by subdividing in the usual binary splitting procedure of cascade, assigning mass $m$ to the left half subinterval and $1-m$ to the right one with probability $p$, and mass $m$ to the right and $1-m$ to the left with probability $1-p$. Then, al realizations have the same singularity spectrum, i.e., the spectrum of the deterministic multifractal corresponding to assigning always mass $m$ to the left. This spectrum has only positive values. The parameter $p$ can be chosen to be any number in $(0, 1)$. Note that this multifractal is a special case of the class considered. The crossover from to deterministic in this example is based on the same essence leading to the transform from Eq.(1) to Eq.(2). From this trivial example, one may finds again that the multifractal description is degenerate at many levels with respect to the physics that generates such measures [20,21]. Hence, two-point statistics of multifractal measures is necessary [9,42,32], rather than one-point statistics.

### 4.3 Falconer's example and Mandelbrot's "trio" multifractals

Consider the example in Ref. [19]. One may find that $w_1(-\infty) = w_7(\infty) = 4$, $f[\alpha(-\infty)] = -2$ and $f[\alpha(\infty)] = -1$. It is easy to see that $p_1 < 1$ and $p_7 < 1$. These results are consistent to the ones via explicit expression of $\tau(q)$. In addition, Falconer presented two special point of $q_{\text{crit}}$ with

$$f[\alpha(q_{crit})] = 0 \tag{28}$$

For unsolvable cases, one has to compute $q_{crit}$ numerically. The solution property of Eq.(28) corresponds to the three situations argued before. The situations of the right part of multifractal function are listed in Table 1. Although Falconer's theorem doesn't concern with negative $f(\alpha)$, he nevertheless presented intuitive results about negative dimensions and argued the interval such that $f(\alpha) \geq 0$. Similarly, Mandelbrot [39] proposed a class of random multinomial measures that he called "trio" multifractals and interpret negative dimensions in the term of large deviation theorem. One can readily find that $w_1(-\infty) = w_3(\infty) = 4$ and $f(\alpha(\pm\infty)) = -2$.

**Table 1.** Three situations of the right part of the $f(\alpha)$ curve

| | | | | |
|---|---|---|---|---|
| Case I | $f[\alpha(-\infty)] > 0$ | $w_i(-\infty) < 1$ | $\sum_{i \in \overline{M}} p_i > 1$ | $q_{crit}$ not exists |
| Case II | $f[\alpha(-\infty)] = 0$ | $w_i(-\infty) = 1$ | $\sum_{i \in \overline{M}} p_i = 1$ | $q_{crit} = -\infty$ |
| Case III | $f[\alpha(-\infty)] < 0$ | $w_i(-\infty) > 1$ | $\sum_{i \in \overline{M}} p_i < 1$ | $-\infty < q_{crit} < 0$ |

*4.4 An "anomalous" random multifractal*

We present an artificial example, which seems anomalous to the usual random multifractals. The construction rule matrix is given by

$$\mathbf{CRM} = \begin{bmatrix} 1/16 & 1/32 & 1/2 & 3/16 & 1/2 & 5/16 & 1/4 \\ \sqrt{2}/4 & 243/1024 & \sqrt{2}/4 & 1/8 & 3/4 & 1/8 & 1/2 \\ 1/1024 & 1/4 & 1/2 & 1/4 & 1/2 & 1/4 & 1/4 \end{bmatrix} \tag{29}$$

It is easy to see that $\overline{M} = \{4,8\}$ and $\underline{M} = \{2,5,6\}$. According to Sec. 3, one finds that $w_4(-\infty) = 0.7869$, $w_8(-\infty) = 0.8523$, and $w_i(\infty) = 1$ for $i \in \underline{M}$. Hence, we have $f(\alpha(-\infty)) = 0.2305$ from Eq. (25) and $f(\alpha(\infty)) = 0$ from Eq. (26).

To illustrate the behavior of each element $w_i(q)$ of MEGF, we evaluate $\tau(q)$ numerically according to Eq.(4) and hence $w_i(q)$ according to Eq.(3). The plots of $w_i(q)$ with $i = 2,5,6$, $i = 4,8$ and $i = 1,3,7$ are shown respectively in Fig. 1a, 1b and 1c, which verify the monotonicity and asymptotic behaviors of $w_i(q)$ argued in Sec. 3. In addition, $\alpha(q)$ are calculated by using forward difference scheme. Then one obtains $f(\alpha)$ via simple algebraic manipulations. The graph of the multifractal spectrum is illustrated in Fig. 1d, which is tangent with the diagonal of the first quadrant $f(\alpha) = \alpha$. This feature implies that the random measure is conservative. The fractal dimension $D_f$ of the random Cantor support is 0.8062 where $q = 0$.

The range of the strength of singularity is from 0.2 to 2, while the multifractal function takes values between 0 and 0.8062. We find that the multifractal function $f(\alpha)$ is nonnegative for every $\alpha \in (\alpha_{\min}, \alpha_{\max})$.

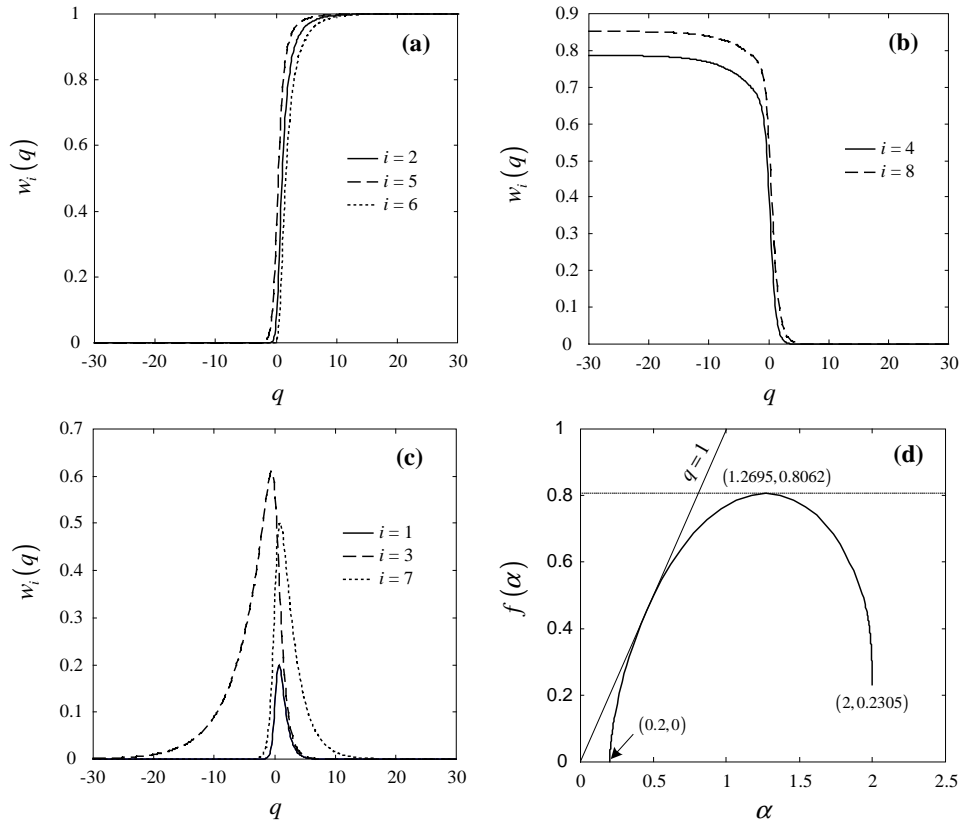

Fig.1 An illustration: (a) Dependence of $w_i(q)$ upon $q$ for $i \in \underline{M}$, (b) Dependence of $w_i(q)$ upon $q$ for $i \in \overline{M}$ and (c) Dependence of $w_i(q)$ upon $q$ for $i \in I - M$. Graph (d) is the multifractal spectrum. The range of the strength of singularity is from 0.2 to 2, while the multifractal function takes values between 0 and 0.8062. That is, the multifractal function is nonnegative.

## 5  Conclusions

Latent dimensions appear usually in random multifractals in both physical systems and mathematical realizations. We studied a class of random multinomial measures and find it is possible that the $f(\alpha)$ function may be always non-negative under proper conditions. We have argued the asymptotic behaviors of the multifractal function $f(\alpha)$ as $q \to \infty$. The so-called latent dimensions condition is presented which states that latent dimensions emerge if and only if the sum of certain probability is less than unity and so negative $f(\alpha)$ may be absent in random multifractal measures.

We investigate several examples in the viewpoint of partition equation and compare the numerical result of simulated realizations with the theory. The agreement is remarkable. It is expected to verify our arguments in real physical systems containing multinomial measures, say, the distribution of voltage drops in a random hierarchical network of resistors [54].

## 6   Acknowledgements

We thank J Barral for providing relevant papers, and especially L Olsen for providing a variety of his reprints and fruitful comments and suggestions on the paper. This work was supported by National Development Programming of Key and Fundamental Researches of China (No. G1999022103).

**References**

1. Arbeiter M and Patzschke N., Random self-similar multifractals. *Math. Nachr.* **181** (1996) pp. 5-42.
2. Barral J., Continuity of the multifractal spectrum of a random statistically self-similar measure. *J. Theor. Prob.* **13** (2000) pp. 1027-1060.
3. Benzi R., Paladin G., Parisi G. and Vulpiani A., On the multifractal nature of fully developed turbulence and chaotic systems. *J. Phys. A* **17** (1984) pp. 3521-3531.
4. Bershadskii A., Critical points of generalized scaling. *Physica A* **268** (1999) 142-148.
5. Biferale L., Boffetta G., Celani A. and Toschi F., Multi-time, multi-scale correlation functions in turbulence and in turbulent models. *Physica D* **127** (1999) pp. 187-197.
6. Blumenfeld R. and Aharrony A., Breakdown of multifractal behavior in diffusion-limited aggregation. *Phys. Rev. Lett.* **62** (1989) pp. 2977-2980.
7. Brachet M. E., Belin F., Tabeling P. and Willaime H., Multifractal asymptotic modeling of the probability density function of velocity increments in turbulence. *Physica D* **129** (1999) pp. 93-114.
8. G. Brown, G. Michon, J. Peyrière, On the multifractal analysis of measure. *J. Stat. Phys.* **66** (1992) pp. 775-790.
9. Cate M. E. and Deutsch J. M., Spatial correlations in multifractals. *Phys. Rev. A* **35** (1987) pp. 4907-4910.
10. Cates M. E. and Witten T. A., Diffusion near absorbing fractals: Harmonic measure exponents for polymers. *Phys. Rev. A* **35** (1987) pp. 1809-1834.

11. Cawley R. and Mauldin R. D., Multifractal decomposition of Moran fractals. *Adv. Math.* **92** (1992) pp. 196-236.
12. Cheng Q. M., The gliding box method for multifractal modeling. *Computers and Geosciences* **25** (1999) pp .1073-1079.
13. Chhabra A. B. and Jensen R. V., Direct determination of the $f(\alpha)$ singularity spectrum. *Phys. Rev. Lett.* **62** (1989) pp. 1327-1330.
14. Chhabra A. B., Meneveau C., Jensen R. V. and Sreenivasan K. R., Direct determination of the $f(\alpha)$ singularity spectrum and its application to fully developed turbulence. *Phys. Rev. A* **40** (1989) pp. 5284-5294.
15. Chhabra A. B. and Sreenivasan K. R., Negative dimensions: theory, computation and experiment. *Phys. Rev. A* **43** (1991) pp. 1114-1117.
16. Chhabra A. B. and Sreenivasan K. R., Scale-invariant multiplier distribution in turbulence. *Phys. Rev. Lett.* **68** (1992) pp. 2762-2765.
17. Chernoff H., A measure of asymptotic efficiency for tests of a hypothesis based on the sum of observations. *Ann. Math. Stat*. **23** (1952) pp. 493-507.
18. Evertsz C. J. G. and Mandelbrot B. B., Multifractal measures. Chaos and Fractals, Peitgen H. O., Jürgens H. and Saupe D. New York: Springer-Verlag, 1992.
19. Falconer K. J., The multifractal spectrum of statistically self-similar measures. *J. Theor. Prob*. **7** (1994) pp. 681-702.
20. Feigenbaum M. J., Some characteristics of strange sets. *J. Stat. Phys*. **46** (1987) pp. 919-924.
21. Feigenbaum M. J., Scaling spetra and return times of dynamical systems. *J. Stat. Phys.* **46** (1987) pp. 925-932.
22. Frisch U., Turbulence: The Legacy of A. N.Kolmogorov, Cambridge University, Cambridge, 1996.
23. Frisch U. and Parisi G., in: Turbulence and Predictability in Geophysical fluid Dynamics, M. Gil, R. Benzi, G. Parisi, Eds., North-Holland, New York, 1985. pp.84-88.
24. Grassberger P. Generalized dimensions of strange attractors. *Phys. Lett. A* **97** (1983) pp. 227-230.
25. Grassberger P., Generalizations of the Hausdorff dimension of fractal measure. Physics Letters A **107** (1985) pp. 101-105.
26. Grassberger P., Badii R. and Politi A., Scaling law for invariant measures on hyperbolic and nonhyperbolic attractors. *J. Stat. Phys.* **51** (1988) pp. 135-177.

27. Grassberger P. and Procaccia I., Dimensions and entropies of strange attractors from a fluctuating dynamics approaches. *Physica D* **13** (1984) pp. 34-54.
28. Halsey T. C., Jensen M. H., Kadanoff L. P., Procaccia I. and Shraiman B. I., Fractal measures and their singularities: the characterization of strange sets. *Phys. Rev. A* **33** (1986) pp. 1141-1151.
29. Hentschel H. G. E. and Procaccia I., The infinite number of generalized dimensions of fractals and strange attractors. *Physica D* **8** (1983) pp. 435-444.
30. Jouault B, Greiner M. and Lipa P., Fix-point multiplier distributions in discrete turbulent cascade models. *Physica D* **136** (2000) pp.125-144.
31. Jouault B, Lipa P. and Greiner M., Multiplier phenomenology in random multiplicative cascade processses. *Phys. Rev. E* **59** (1999) pp. 2451-1454.
32. Lee S. J. and Halsey T. C., Some results on multifractal correlations. *Physica A* **164** (1990) pp. 575-592.
33. Lee J. and Stanley H. E., Phase transition in the multifractal spectrum of diffusion-limited aggregation. *Phys. Rev. Lett.* **61** (1988) pp. 2945-2948.
34. Lepreti F., Carbone V., Pietropaolo E., Consolini G., Bruno R., Bavassano B. and Berrilli F., Multifractal structure of the dissipation field of intensity fluctuations in the solar photosphere. *Physica A* **280** (2000) pp. 80-91.
35. Mandelbrot B. B., Intermittent turbulence in self-similar cascade: divergence of high moments and dimension of carrier. *J. Fluid Mech.* **62** (1974) pp. 331-358.
36. Mandelbrot B. B., Multifractal measures, especially for the geophysicist. *Pure Appl. Geophys*. **131** (1989) pp. 5-42.
37. Mandelbrot B. B., Limit lognormal multifractal measures. in: Frontiers of Physics: Landau Memorial Conference, E. Dotsman, Y. Ne'eman, A. Voronel, eds., Pergamon, New York, 1989.
38. Mandelbrot B. B., A class of multinomial multifractal measures with negative (latent) values for the "dimension" $f(\alpha)$. in: Fractals' Physical Origin and Properties, L. Pietronero, ed., Plenum, New York, 1989. pp. 3-29.
39. Mandelbrot B. B., Negative fractal dimensions and multifractals. *Physica A* **163** (1990) pp. 306-315.
40. Mandelbrot B. B., Random multifractals: negative dimensions and the resulting limitations of the thermodynamic formalism. *Proc. Roy. Soc. London A* **434** (1991).

41. Meatin P., Coniglio A., Stanley H. E. and Witten T. A., Scaling properties for the surfaces of fractal and nonfractal objects: An infinite hierarchy of critical exponents. *Phys. Rev. A* **34** (1986) 3325-3340.
42. Meneveau C. and Chhabra A. B., Two-point statistics of multifractal measures. *Physica A* **164** (1990) pp. 564-574.
43. Meneveau C. and Sreenivasan K. R., Measurement of $f(\alpha)$ from scaling of histograms, and applications to dynamical systems and fully developed turbulence. *Phys. Lett. A* **137** (1989) pp. 103-112.
44. Meneveau C. and Sreenivasan K. R., The multifractal nature of turbulent energy dissipation. *J. Fluid Mech.* **224** (1991) pp. 429-484.
45. Olsen L., Multifractal formalism. *Adv. Math*. **116** (1995) pp. 92-195.
46. Olsen L., Multifractal slices and negative dimensions, Preprint, 1998.
47. Olsen L., Geometric constructions in multifractal geometry. *Periodica Methematica Hungaria* **37** (1998) pp. 81-99.
48. Olsen L., Measurability of multifractal measure functions and multifractal dimension functions. *Hiroshima Math. J.* **29** (1999) pp. 435-458.
49. Olsen L., Multifractal geometry. *Progress in Probability* **46** (2000) pp. 3-37.
50. Prasad R. R., Meneveau C. and Sreenivasan K. R., Multifractal nature of the dissipation field of passive scalars in fully turbulent flow. *Phys. Rev. Lett.* **61** (1988) pp. 74-77.
51. Prasad R. R. and Sreenivasan K. R., Quantitative three-dimensional imaging and the structure of passive scalar fields in fully developed turbulent flow. *J. Fluid Mech.* **216** (1990) pp. 1-34.
52. Schwarzer S., Lee J., Bunde A., Halvin S., Roman H. E and Stanley H. E., Minimum growth probability of diffusion-limited aggregation. *Phys. Rev. Lett*. **65** (1990) pp. 603-606.
53. Sreenivasan K. R. and Meneveau C., The fractal facets of turbulence. *J. Fluid Mech.* **173** (1986) pp. 357-386.
54. Vicsek T., Fractal Growth Phenomena, World Scientific, Singapore, 1992.
55. Zhou W. X. and Yu Z. H., Multifractality of drop breakup in the air-blast nozzle atomization process. *Phys. Rev. E* **63** (2001) pp. 016202.